\documentclass[aps,a4paper,10pt,twocolumn,nofootinbib]{revtex4} 
\usepackage[T1]{fontenc}
\usepackage{mathptmx}
\usepackage{datetime}
\usepackage{amsmath}
\usepackage{amsfonts}
\usepackage{amsfonts}
\usepackage{mathrsfs}
\usepackage[mathscr]{euscript}
\usepackage[dvips]{graphicx}
\usepackage{fancyhdr}
\usepackage{color}
\usepackage{hyperref}
\usepackage{epsfig}
\usepackage{color}
\usepackage{bm}

\usepackage{tikz}

\def\hwfrac{0.75*}
\def\minthick{0.33}

\definecolor{Blue}{rgb}{0.1,0.1,0.85}
\definecolor{Red}{rgb}{0.85,0.1,0.1}

\def\overstrike#1#2{{\setbox0\hbox{$#2$}\hbox to \wd0{\hss
    $#1$\hss}\kern-\wd0\box0}}

\renewcommand{\Vec}{\bm}

\newdateformat{yymmdddate}{\THEYEAR/\twodigit{\THEMONTH}/\twodigit{\THEDAY}}

\newcommand{\XDOI}[1]{\href{http://dx.doi.org/#1}{doi:#1}}
\newcommand{\XARXIV}[1]{\href{http://arxiv.org/abs/#1}{arXiv:#1}}
\newcommand{\XWEB}[1]{\href{#1}{#1}}

\newcommand{\pTime}{t}

\newcommand{\pSpaceX}{x}

\newcommand{\pSpaceZ}{z}

\newcommand{\pWaveLength}{\lambda}          

\newcommand{\pFreqc}{f}                     
\newcommand{\pFrang}{\omega}                
\newcommand{\pWaveVectorS}{k}          

\newcommand{\pPderivT}{\partial_{\pTime}}   
\newcommand{\pPderivX}{\partial_{\pSpaceX}} 
\newcommand{\pPderivZ}{\partial_{\pSpaceZ}} 

\newcommand{\pPermittivity}{\epsilon}      
\newcommand{\pPermittivityVac}{\epsilon_0} 

\newcommand{\pPermeabilityVac}{\mu_0}      

\newcommand{\pXemXelectric}{E}                      
\newcommand{\pXemXelectricv}{\Vec{\pXemXelectric}}  
\newcommand{\pXemXpolarization}{P}                         
\newcommand{\pXemXpolarizationv}{\Vec{\pXemXpolarization}} 

\newcommand{\pXemXspotential}{\phi}        

\newcommand{\pCurrent}{J}                  
\newcommand{\pCurrentv}{\Vec{\pCurrent}}   

\newcommand{\pEfield}{\pXemXelectric}

\newcommand{\pEfieldv}{\pXemXelectricv}

\def\Dmarker{\Large{$\times$}}
\def\tblue{blue!80!black!99}

\def\cBlue#1{{\color{blue}{#1}}}
\def\cBlue#1{#1}

\begin{document}
\title{An introduction to spatial dispersion: \\
revisiting the basic concepts}
\author{Paul Kinsler}
\homepage[]{https://orcid.org/0000-0001-5744-8146}
\email[\hphantom{.}~]{Dr.Paul.Kinsler@physics.org}
\affiliation{
  Cockcroft Institute,
  Keckwick Lane, Daresbury WA4 4AD, 
  United Kingdom
}
\affiliation{
  Physics Department,
  Lancaster University, 
  Lancaster LA1 4YB,
  United Kingdom
}
\address{
  Department of Physics,
  Imperial College London,
  London SW7 2AZ,
  United Kingdom
}

\lhead{\includegraphics[height=5mm,angle=0]{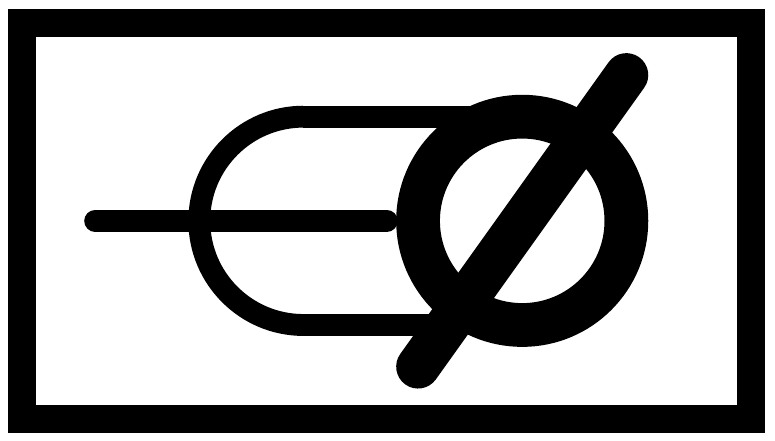}~~SPATYPE}
\chead{An introduction to spatial dispersion}
\rhead{
\href{mailto:Dr.Paul.Kinsler@physics.org}{Dr.Paul.Kinsler@physics.org}\\
\href{http://www.kinsler.org/physics/}{http://www.kinsler.org/physics/}
}

\begin{abstract}

I describe three ways that the spatial properties
 of a wave propagation medium
 can cause dispersion, 
 and propose that they should form the basics for
 correctly understanding and \emph{naming}
 phenomena described as ``spatial dispersion''.
In particular, 
 I emphasise the specific spatial properties which 
 generate the resulting dispersive --
 i.e. \emph{spatially dispersive} --
 behaviour.
The properties are
 geometry, 
 structure, 
 and
 dynamics.

\end{abstract}


\date{\today}
\maketitle
\thispagestyle{fancy}

%

%
\section{Introduction}\label{S-intro}

Spatial dispersion is the non-local dependence
 of material properties on both direction and wavelength,
 and is of particular interest in electromagnetic systems
 and other fields where artificial functional materials (AFMs)
 are finding technological uses, 
 such as in acoustic and elastic metamaterials.
It has been long observable in crystals \cite{AgranoGinsberg}, 
 but is also of particular importance in AFMs
 where the wavelength of radiation
 becomes comparable to the lattice parameters 
 (see e.g. \cite{Belov-MMNSST-2003prb}).
Further, 
 it is also significant in the region of 
 the material resonances utilized in metamaterial unit cells,
 where constitutive parameters such as 
 the permittivity and permeability --
 and likewise their acoustic or elastic counterparts --
 are maximised or have near-zero \emph{local} values 
 \cite{Pollard-MHEAWZP-2009prl}
 or where the response is non-reciprocal \cite{Xu-GHLLC-2013nc}.
In some materials its presence is revealed 
 by the shape of the equi-frequency surfaces of the dispersion relations --
 e.g. when they are non-symmetric,
 not elliptical or hyperbolic,
 or having multiple modes with the same direction and polarisation.


Spatial dispersion
 usually appears
 as a either spatially non-local effect
 that produces
 a wavevector dependence of the material parameters,
 or
 as a non-trival wavevector dependence for the dispersion relations.
{Since representing spatial properties in a fourier space
 (here,
 that of wavevector $\pWaveVectorS$ )
 necessarily involves an integral over spatially separated points, 
 spatial dispersion is intrinsically non-local.}

Although spatial dispersion can be modelled
 in an ad hoc manner
 to suit intuition or an empirical fit to data, 
 it is preferable to be more systematic.
However, 
 the term ``spatial dispersion'' is often used rather loosely,
 and so sometimes the origin
 of the specific phenomenon being discussed
 is unclear.
In an attempt to rectify this, 
 here I describe three physically distinct ways that 
 spatial dispersion can arise, 
 briefly describing examples of each.
Each of these mechanisms has 
 (a) no time dependence beyond that required to support a wave, 
 and 
 (b) a specification of time-independent  material properties.
The lack of any (non-trivial) time dependence 
 ensures that the phenomena treated here are entirely unrelated
 to the more commonly considered topic of temporal dispersion.
The first two spatial dispersion mechanisms
 {depend on inhomogeneous spatial properties, 
 whereas the third derives from 
 propagation-based properties.}

The categorization here shows that instances of spatial dispersion
 can be caused by the
 geometric, 
 structural,
 or dynamic properties of a system.
It is not based on existing
 (and often somewhat ad hoc)
 naming conventions or justifications for spatial dispersion,
 but is instead grounded in the following assertion:
~\\

\noindent
\textbf{\textit{Spatial dispersion refers to any dispersive behaviour
 that occurs (solely) as a result of 
 spatial properties of the system.}}\\
~

This viewpoint is distinct from a widely used one
 (see e.g. \cite{Klingshirn-SO})
 where spatial dispersion is 
 characterised as being due to (non-trivial) 
 wavevector-dependence in the dispersion relations,
 and is then 
 categorized into two cases
 `weak' spatial dispersion,
 where 
 the spatial effects have been approximated as a local effect;
 and `strong' spatial dispersion, 
 where nonlocal effects remain important
 \cite{Khrabustovskyi-MPSR-2017arxiv,Mnasri-KSPR-2018prb}.
It is to be noted that so-called weak spatial dispersion
 is not necessarily small in any practical sense, 
 despite the implications of `weak'; 
 and likewise so-called strong spatial dispersion
 need not be particularly dominant\footnote{In 
  my view, 
  it is preferrable for
  terms such as `weak' or `strong' to be only used to 
  to indicate the strength or significance of a phenomenon,
  not its origin or type,
  or whether it might be effectively local or non-local.}.

~\\
In particular, 
 in the interest of both brevity
 and not repeating existing discussions in the literature,
 I do not:

\begin{enumerate}

\item
discuss how homogenization schemes might 
 reduce the description of a complicated medium
  {--
 such as those with spatial varying \emph{temporal} responses --}
 into a simpler one, 
 (e.g.) perhaps one based on a series expansion
 in terms of the wavevector $\pWaveVectorS$;

\item
 consider dispersion relations
 that have been artificially constructed 
 \emph{without} reference
 to some specific physical system with spatial properties;
 e.g. putative dispersia based on
 suggested or ``what-if'' polynomials in frequency $\pFrang$
 and wavevector $\pWaveVectorS$,
 some other specification of an $\pFrang$, $\pWaveVectorS$ interdependence,
 or a non-local convolution over spatial properties.

\end{enumerate}

The main focus here 
 is on spatial dispersion in electromagnetic systems
 or artifical functional media such as metamaterials, 
 but the distinctions apply equally well
 to waves in acoustic or elastic materials, 
 or indeed potentially \emph{any} material system
 which supports wave propagation of some kind.
In Section \ref{S-dispersion},
 I will define
 what I mean by dispersion
 (i.e. specifically non-trivial dispersion)
 by considering the relationships between 
 the frequency $\pFrang$ of a wave 
 and the wavevector $\pWaveVectorS$
 associated with that frequency.
Following that the next three sections 
 address each of 
 geometric (Section \ref{S-Geometric}), 
 structural (\ref{S-Structural}),
 and
 dynamic (\ref{S-Dynamic}) 
 spatial dispersions in turn.
Note that the idea of spatial dispersion
 as being grounded in spatial properties 
 as opposed to (spectral) wavevector ones
 is mirrored in the various treatments of temporal dispersion:
 although sometimes seen
 as a consequence of a dynamic time-domain process
 \cite{Kinsler-2011ejp}
 particularly when
 implemented in FDTD algorithms \cite{Oskooi-RIBJJ-2010cpc}, 
 temporal dispersion
 is more often discussed in solely terms of a (spectral) frequency response
 \cite{Toll-1956pr,Dirdal-BS-2015prb}.
{Of course, 
 in practical terms
 the categorization of dispersion types is really applied to
 our models or representations of the physical system, 
 and not some perfectly accurate microscopic view of
 the actual physical system, 
 and this is discussed in Section \ref{S-discussion}.}

In what follows, 
 I do not 
 describe the example systems in a detailed manner.
Only the minimum features necessary to make the point are presented,
 as such models are often worked through 
 both exhaustively and frequently
 in other sources.
Here their role is simply to provide examples 
 that typify their role in creating a dispersive response 
 as a result of their particular spatial features or structure.
Nonlinear effects are not considered.

\section{Dispersion}\label{S-dispersion}

\def\waveindex{m}

The definition of dispersion 
 used here is based on three criteria
 relating to how the frequency ($\pFreqc$ or $\pFrang$) of a wave
 is related to its wavelength $\pWaveLength$ or wavevector $\pWaveVectorS$.
Note that 
 in discrete systems, 
 analytic solutions typically depend on an integer index $\waveindex$
 (and possibly more indices as well).
Since these indices usually indicate the number
 or spatial frequency of the oscillation in a given solution, 
 they play the same role as the wavevector $\pWaveVectorS$, 
 and contain essentially the same physical content.
In simple cases, 
 the wavevector $\pWaveVectorS$
 can be a straightforward multiple of $\waveindex$; 
 in any case here the wavevector associated with 
 an index $\waveindex$ is $\pWaveVectorS_{\waveindex}$,
 and likewise the related angular frequency is $\pFrang_{\waveindex}$.

The criteria are:

\begin{description}

\item[Linear]--
 the relationship between $\pFrang$ and $\pWaveVectorS$ 
 (or between $\pFrang$ and $\waveindex$ or $\pWaveVectorS_{\waveindex}$) 
 is linear, 
 i.e. it lies along a single straight line.
 \label{criteria-linear}

\item[Origin]--
 if $\pFrang=0$ then $\pWaveVectorS=0$
 (or $\waveindex=0$), 
 and vice versa.
 \label{criteria-zero}

\item[Continuous]--
 the relationship is continuous,
  i.e. 
  is in one piece, 
  and has no gaps or jumps.
 \label{criteria-continuous}

\end{description}

\begin{figure}
{\centering
\resizebox{0.70\columnwidth}{!}{
\begin{tikzpicture}
\draw [line width=2,<->] (0,4) -- (0,0) -- (4,0) ;
\draw (-0.3,3) node {$\pFrang$} ;
\draw (3,-0.3) node {$\pWaveVectorS$} ;
\draw [line width=1, \tblue] (0,0) -- (3.2, \hwfrac 3.2) ;
\draw [line width=\minthick, dotted, \tblue] (3.2, \hwfrac 3.2) -- (3.8, \hwfrac 3.8) ;
\end{tikzpicture}
}}
\caption{Dispersionless case, 
 where the dispersion relation is linear, 
 passes through and includes the origin, 
 and is continuous.
The light dotted line indicates
 that the depicted dispersion behaviour continues
 in a similar manner 
 for higher wavevectors and frequencies.}
\label{fig-displess}
\end{figure}
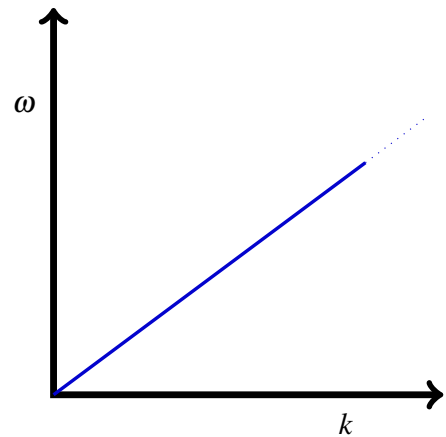

If all these criteria hold, 
 then the wave is \emph{dispersionless}, 
 and the 
 phase and group velocities are always the same.
If any of the criteria do not hold, 
 then the wave is dispersive,
 and an initial pulse shape will change over time 
 as its different spectral components evolve at different rates.
In general,
 this behaviour 
 might be due to either the time-response of the medium
 (giving rise to temporal dispersion), 
 or to the spatial properties of the medium
 (giving rise to spatial dispersion), 
 or possibly both.

Here I focus exclusively on dispersion
 that results solely from \emph{spatial} properties.
However, 
 it is worth keeping in mind that 
 typically the structural elements
 relied on in a physical device to create those spatial properties 
 may in reality also be temporally dispersive --
 notably,
 reflecting walls will not be perfectly reflective at all frequencies, 
 and
 different refractive indices exist primarily because of temporal dispersion;
 some of these issues are discussed in section \ref{S-discussion}.
In what follows I ignore such complications 
 because I wish to focus solely on spatial properties and their effects; 
 but this is not to say that such time-domain responses may always be ignored
 in practical situations.

A dispersionless wave
 travelling along the $z$--axis
 with propagation speed $c$
 follows the wave equation
~
\begin{align}
  \pPderivT^2 \pEfield
 -
  \pPderivZ c^2 \pPderivZ \pEfield
&=
  0
,
\label{eqn-dispersionless-tz}
\end{align}
 where $\pPderivT \equiv d/dt$, 
 and $\pPderivZ \equiv d/dz$.
This equation can be Fourier transformed in both time and space, 
 and the field strength terms cancelled,
 giving the ``dispersionless'' dispersion relation
~
\begin{align}
  \pFrang^2 - c^2 \pWaveVectorS^2 = 0
.
\label{eqn-dispersionless-wk}
\end{align}
A depiction of this 
 relationship between $\pFrang$ and $\pWaveVectorS$
 can be seen on fig. \ref{fig-displess}.

Note that this is quadratic in both $\pFrang$ and $\pWaveVectorS$, 
 so that either might have negative values and still help satisfy 
 the dispersion relation.
Any physical meaning(s) that is attributable
 to such negative spectral quantities is 
 comprehensively discussed elsewhere
 \cite{Kinsler-2014arXiv-negfreq}, 
 so for simplicity we will only consider positive values here.
Typically the sign choices correspond to the different direction
 in which propagating waveforms will evolve
 \cite{Kinsler-2010pra-fchhg,Kinsler-2018jo-d2owe,Kinsler-2018jpco-fbacou}.

%
\section{Geometric spatial dispersion}\label{S-Geometric}

The \emph{geometry} of a region 
 in which the wave of interest is supported
 is perhaps the most primitive possible source 
 of a spatial effects like dispersion.

For the purposes here, 
 I consider 
 a (purely) geometric system to be one 
 where the properties of the propagation medium
 are homogeneous,
 isotropic, 
 and are not temporally dispersive.
This medium is present everywhere
 except at any boundaries,
 and to guarantee complete confinement within the propagation medium,
 boundaries are assumed to be 
 perfectly reflective.  
Note that this category also includes systems without boundaries, 
 such as those confined on (e.g.) 
 a closed surface such as 
 a torus
 or a sphere.
{In confined (cavity) systems, 
 the non-local property causing the spatial dispersion
 is the separations of those boundaries, 
 in boundary-free systems it is the size.
Although any one point on a boundary is local, 
 separated points are not.}

In such systems, 
 especially if there is sufficient symmetry, 
 we can often find analytic solutions for its frequency eigenmodes.
In such a case, 
 we have an index ${\waveindex}$
 rather than a Fourier-transform based wavevector, 
 but the role is the same:
 ${\waveindex}$ indicates the number of spatial oscillations
 over some relevant distance interval;
 typically it is closely related to the number of nodes
 in its eigenfunction.

%
\subsection{Cavity}\label{S-Geometric-cavity}

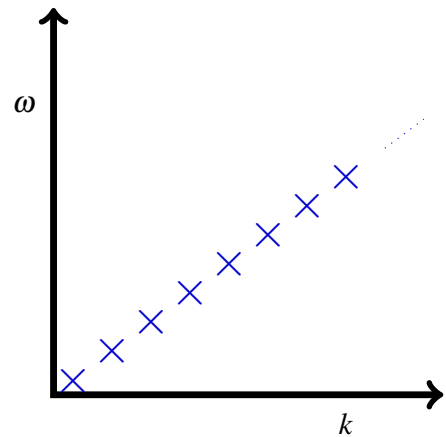
\begin{figure}
{\centering
\resizebox{0.70\columnwidth}{!}{
\begin{tikzpicture}
\draw [line width=2, <->] (0,4) -- (0,0) -- (4,0) ;
\draw (-0.3,3) node {$\pFrang$} ;
\draw (3,-0.3) node {$\pWaveVectorS$} ;
\foreach \i in {0.2,0.6,...,3.4} do {
 \draw [\tblue] (\i, \hwfrac \i) node {\Dmarker} ;
} ;
\draw [line width=\minthick, dotted, \tblue] (3.4, \hwfrac 3.4) -- (3.8, \hwfrac 3.8) ;
\end{tikzpicture}
}}
\caption{Geometric spatial dispersion in a 1D cavity --
 the dispersion relation
 in this case is a series of points, 
 as indicated by the crosses.
It is linear, 
 does not include the origin, 
 and is not continuous.}
\label{fig-geom-cavity}
\end{figure}

Although any shape of wave-confining, 
 empty cavity would a candidate 
 for supporting spatial dispersion,  
 the simplest would be
 a 1D perfectly reflective 
 cavity with a length $L$.
In such a case,
 where the wave amplitude is zero on the boundary,
 the dispersion relation is 
~
\begin{align}
  \pFrang = \frac{{\waveindex}+1}{2} \frac{2\pi c}{L}
,
\end{align}
 where ${\waveindex}$ is a non-negative integer, 
 and the equivalent wavevector is 
 $\pWaveVectorS_{\waveindex} = \pi ({\waveindex}+1) / L$.
The dispersion relation for this system 
 is depicted schematically on fig. \ref{fig-geom-cavity}.

By the criteria set in Section \ref{S-dispersion},
 this \emph{is} spatially dispersive since 
 although the relationship between $\pFrang$ and $\pWaveVectorS$
 is linear,
 with $\pFrang = c \pWaveVectorS$,
 there is no supported wave with $\pFrang = 0$ and  $\pWaveVectorS = 0$, 
 and only
 a discrete spectrum of waves exist.

The primary effect of spatial dispersion here 
 is simply the imposition of a discrete spectrum, 
 although the removal of any 
 $(\pFrang,\pWaveVectorS) = (0,0)$ solution is also important.
Other cavity shapes, 
 such as cylindrical or spherical, 
 are widely covered in undergraduate textbooks on electromagnetism
 so I do not present them here, 
 but they also will (at least)
 have a discrete spectrum and no $(0,0)$ solution.

%
\subsection{Topology: torus}\label{S-Geometric-topologyA}

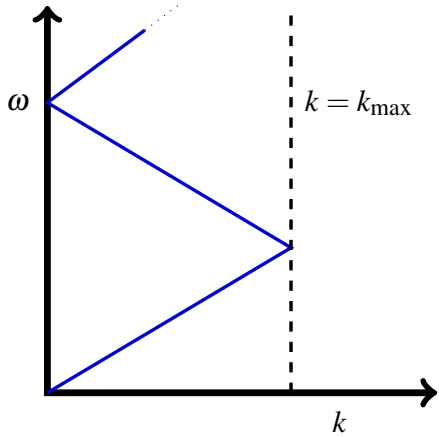
\begin{figure}
{\centering
\resizebox{0.70\columnwidth}{!}{
\begin{tikzpicture}
\draw [line width=2, <->] (0,4) -- (0,0) -- (4,0) ;
\draw (-0.3,3) node {$\pFrang$} ;
\draw (3,-0.3) node {$\pWaveVectorS$} ;
\draw [line width=1, \tblue] (0,0) -- (2.5, \hwfrac 2) -- (0, \hwfrac 4) -- (1.0, \hwfrac 5.0);
\draw [line width=\minthick, dotted, \tblue] (1.0, \hwfrac 5.0) -- (1.35, \hwfrac 5.35) ;
\draw [line width=1, dashed] (2.5,0) -- (2.5,4) ;
\draw (3.2,3) node {$k=k_\textup{max}$} ;
\end{tikzpicture}
}}
\caption{Geometric spatial dispersion on a loop (a 1D torus) --
 the dispersion relation is piece-wise linear
 but not strictly linear, 
 does include the origin, 
 and is continuous.}
\label{fig-geom-torus}
\end{figure}

A wave-supporting space without boundaries 
 is a candidate for supporting spatial dispersion
 (only) if it is also finite, 
 so that it's size provides an intrinsic length scale.
The simplest situation is probably a toroidal space,
 which is essentially the same as the case of
 an infinite and periodic lattice, 
 or indeed of periodic boundary conditions.
In 1D, 
 the torus is a simple loop,
 with the length $L$ giving the periodicity scale 
 and setting an effective
 maximum wavevector $\pWaveVectorS_{\textup{max}} = 2\pi/L$.
In a dispersion plot, 
 this gives rise to band folding \cite{joannopoulos2011photonic} --
 i.e. although a non-periodic dispersion relation would normally extend
 to both high wavevector and high frequency,
 those parts of the dispersion at too-high wavevector
 are ``folded'' back to lower wavevector; 
 so that
 the periodic case is restricted to a finite $\pWaveVectorS$ range,  
 as can be seen on fig. \ref{fig-geom-torus}.
This means that at low wavevectors and lowest frequencies, 
 the system appears dispersionless, 
 but once band folding occurs for the higher frequencies,
 even though the dispersion might be linear and continuous,
 the folded dispersion can no longer be extrapolated
 from every point
 to pass through $(0,0)$.

%
\subsection{Topology: sphere}\label{S-Geometric-topologyB}

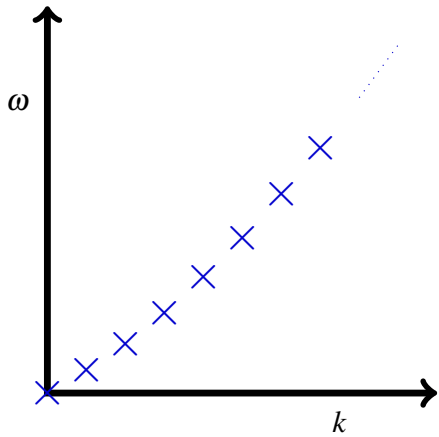
\begin{figure}
{\centering
\resizebox{0.70\columnwidth}{!}{
\begin{tikzpicture}
\draw [line width=2, <->] (0,4) -- (0,0) -- (4,0) ;
\draw (-0.3,3) node {$\pFrang$} ;
\draw (3,-0.3) node {$\pWaveVectorS$} ;
\draw [line width=\minthick, dotted, \tblue]  ( 0.4*8, {\hwfrac 0.16 *sqrt{ 8*9}} )
                       --
                ( 0.4*9, {\hwfrac 0.16 *sqrt{ 9*10}} );
\foreach \i in {0,1,...,7} do {
 \draw [\tblue] ( 0.4*\i, {\hwfrac 0.16 *sqrt{ \i*(\i+1)}} ) node {\Dmarker} ;
} ;
\end{tikzpicture}
}}
\caption{Geometric spatial dispersion on a sphere --
 the dispersion relation
 in this case is a series of points, 
 as indicated by the crosses.
It is nonlinear, 
 includes the origin, 
 but is not continuous.}
\label{fig-geom-sphere}
\end{figure}

A more interesting case than those above  is 
 a resonator where the waves are confined on a spherical surface,
 which is equivalent to a Maxwell's fisheye lens \cite{Luneberg-MTO}.
The supported modes
 or this 
 are derived from the 
 Legendre polynomials\footnote{See \url{http://dlmf.nist.gov/14}} 
 $P_{\waveindex}(\xi)$.
These mode functions $P_{\waveindex}$
 provide a complete and countable orthonormal basis set
 for all possible radial field patterns in the resonator, 
 with 
 the argument $\xi = (r^2-1)/(r^2+1)$ being derived from
 the radial displacement $r$
 from some  choice of preferred origin.
Of course there can also be an angular dependence to field patterns, 
 which can easily be included using the usual angular mode functions 
 \cite{WfmMathWorld-SphericalHarmonic}, 
 but I omit those details here in the interests of brevity.
Each discrete mode has a frequency $\pFrang_{\waveindex}$ determining
 its physical properties as determined from its index ${\waveindex}$, 
~
\begin{align}
  \pFrang_{\waveindex}^2 
\propto
    {\waveindex}\left({\waveindex}+1\right)
\label{eqn-k-dispersion}
\end{align}
 with modes of larger ${\waveindex}$ (or $\pFrang_{\waveindex}$)
 having more spatial oscillations.
A depiction of the spatial dispersion properties on a sphere
 can be seen on fig. \ref{fig-geom-sphere}.

Here the effect of spatial dispersion is significant, 
 because two of the criteria are violated --
 the relationship is not linear,
 and the spectrum is discrete.


%
\section{Structural spatial dispersion}\label{S-Structural}

Structural spatial dispersion is 
 distinct from the 
 geometric 
 spatial dispersion above, 
 in that it is due to material inhomogeneity:
 i,e,
 that there are two or more types of material
 supporting the wave field that are 
 arranged in a structure.
This type is the origin of most instances of spatial dispersion
 that are considered.
As we see in the slab waveguide example below,
 even very simple types of inhomogeneity
 can generate spatial dispersion; 
 however, 
 the necessary calculations 
 are usually non-trivial.
{In these systems the non-local nature is clear --
 the spatial dispersion is due to the arrangements and separations
 of all the different material properties within system.}

%
\subsection{Slab waveguide}\label{S-Structural-slab}

An electromagnetic slab waveguide is one of the simplest structures 
 that might be considered when looking for an example of 
 structural spatial dispersion
 in a non-periodic system.
It consists of a planar slab of one type of material
 sandwiched on either side by half-infinite regions of an alternative material,
 where the difference in material properties allows
 modes to exists that are localised in a way 
 centred on the slab.

Electromagnetic slab waveguides are treated in a wide range of textbooks
 (see e.g. \cite{TG-FiberOptEss},
 or the abbreviated summary in \cite{Kinsler-2018jo-d2owe}
 with a discussion of dispersion handling).
Here the waveguide is taken to have
 thickness $d$ in the perpendicular $x$ direction, 
 propagation in the $z$ direction, 
 and with core and cladding permittivities
 ${\pPermittivity}_1$ and ${\pPermittivity}_2$.
To solve this system, 
 we take Maxwell's equations in component form, 
 assume plane-wave like behaviour in orientations parallel to the slab, 
 and,
 at the boundaries, 
 match the $\sin()$ or $\cos()$ functional form in the core
 with decaying exponentials in the cladding.
Even for this simple slab design,
 the boundary conditions give
 the dispersion relation 
 a non trivial form.
Notably, 
 for the bound modes it is given by the solution to 
 a transcendental equation \cite{TG-FiberOptEss}.
Typically, 
 this is written in a way implying we want to calculate $k_x$ and $\beta=k_z$
 from a specified $\omega$; 
 although a compelling argument can be made \cite{Kinsler-2018jo-d2owe}
 that it is better to calculate $\omega$ from a provided $k_z$.

For the transverse electric (TE) field modes, 
 the traditional presentation shows that
 in a slab waveguide we have
~
\begin{align}
  T(k_x d) 
&=
  k_x^{-1} 
  \sqrt{\omega^2 {\pPermeabilityVac} 
    \left(
      {\pPermittivity}_1 - {\pPermittivity}_2 
    \right)
   -
    k_x^2
  }
,\label{eq-example-zeq}
\\
\textrm{where} \quad
  \beta^2(\omega) 
&=
  \omega^2 {\pPermeabilityVac} {\pPermittivity}_1 - k_x^2
,
\label{eq-example-zk}
\end{align}
 and $T(k_x d)$ is either $+\tan(k_x d)$ or $-\cot(k_x d)$.

Fig. \ref{fig-struct-slab} indicates the appearance
 of the dispersion relations
 for the first few bound modes of such a waveguide.
This slab waveguide case is a somewhat 
similar problem to finding the modes of an optical fibre,
 although the cylindrical symmetry of a fibre 
 means that the solutions involve matching Bessel functions
 across the slab boundaries.

\def\swv{0.80}
\def\swh{0.90}

\begin{figure}
{\centering
\begin{tikzpicture}
\draw [line width=2, <->] (0,4) -- (0,0) -- (4,0) ;
\draw (-0.3,3) node {$\pFrang$} ;
\draw (3,-0.3) node {$\pWaveVectorS$} ;
\draw [line width=\minthick, dotted, \tblue] (4.00*\swh,4.05*\swv) --  (4.30*\swh,4.35*\swv) ;
\draw[line width=1,\tblue] plot [smooth,tension=0.4]
 coordinates {(0.30,0.350*\swv) (0.60*\swh,0.90*\swv) (2.30*\swh,2.40*\swv) (4.00*\swh,4.05*\swv)};
\draw [line width=\minthick, dotted, \tblue] (4.00*\swh,4.20*\swv) --  (4.30*\swh,4.50*\swv) ;
\draw[line width=1,\tblue] plot [smooth,tension=0.4]
 coordinates {(0.60*\swh,1.150*\swv) (0.90*\swh,1.70*\swv) (2.90*\swh,3.30*\swv) (4.00*\swh,4.20*\swv)};
\draw [line width=\minthick,dotted, \tblue] (4.00*\swh,4.35*\swv) --  (4.30*\swh,4.65*\swv) ;
\draw[line width=1,\tblue] plot [smooth,tension=0.4]
 coordinates {(0.90*\swh,1.950*\swv) (1.20*\swh,2.50*\swv) (3.50*\swh,4.00*\swv) (4.00*\swh,4.35*\swv)};
\end{tikzpicture}
}
\caption{Structural spatial dispersion in a slab waveguide --
 the dispersion relation is nonlinear,
 does not include the origin, 
 and is not continuous.
This schematic depiction
 shows exaggerated dispersion curves for the first few bound modes. 
}
\label{fig-struct-slab}
\end{figure}
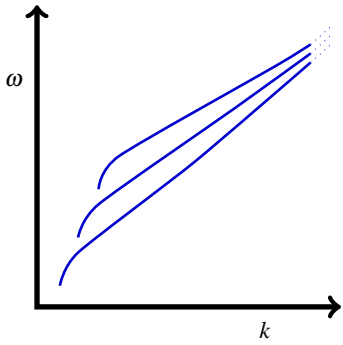

As a passing note, 
 the spatial dispersion due to the structural configuration 
 of an optical fibre
 is often treated as if it were
 instead \emph{temporal} in origin.
This is because a majority of optical pulse propagation techniques
 are propagated along a spatial axis
 \cite{Brabec-K-1997prl,Kolesik-MM-2002prl,Tyrrell-KN-2005jmo,Genty-KKD-2007oe,Kinsler-2010pra-fchhg,Kinsler-2010pra-dblnlGpm,Kinsler-2014arXiv-negfreq}, 
 which means that the dispersion computations are more convenient in $\pFrang$
 than they would be in the $\pWaveVectorS$
 available in temporally propagated techniques
 \cite{Kinsler-2018jo-d2owe,Kinsler-2018jpco-fbacou}.

%
\subsection{Bragg mirror}\label{S-Structural-bragg}

A Bragg mirror is composed of multiple thin layers of dielectric material, 
 with the layers designed so that the device is highly reflective
 at some wavelength, 
 or in some wavelength range.
Simple versions
 consist of stacks of alternating high and low refractive material,
 with thicknesses chosen so that the path-length differences 
 for internal reflections are integer multiples of the design wavelength.
Here, 
 a unit cell for the stack consists of just the two contrasting layers.

In the model calculation of Horsley et al. \cite{Horsley-WAL-2014ajp},
 an infinite periodic stack is considered.
The analysis shows that 
 the relationship between structure and eigenvalue --
 i.e. its dispersion relation is controlled by the expression
~
\begin{align}
  \lambda 
&= 
  z(\omega) \pm \sqrt{z(\omega)^2-1} 
=
  \exp \left[ \imath \kappa \left(a+b\right) \right]
.
\end{align}
Here $\kappa$ is the unit-cell Bloch wavevector, 
 and the layer thicknesses are $a$ and $b$.
The effect of the material refractive indices $n_a$ and $n_b$
 are subsumed into the real valued $z(\omega)$, 
 which is calculated from the unit-cell's transfer matrix.

\begin{figure}
{\centering
\resizebox{0.70\columnwidth}{!}{
\begin{tikzpicture}
\draw [line width=2, <->] (0,4) -- (0,0) -- (4,0) ;
\draw (-0.3,3) node {$\pFrang$} ;
\draw (3,-0.3) node {$\pWaveVectorS$} ;
\draw [line width=1, \tblue] (0,0)              -- (1.25, \hwfrac 1.00) -- (2.40, \hwfrac 1.88) -- (2.5, \hwfrac 1.9);
\draw [line width=1, \tblue] (2.5, \hwfrac 2.1) -- (2.40, \hwfrac 2.12) -- (1.25, \hwfrac 3.00) -- (0.099,   \hwfrac 3.88) -- (0,   \hwfrac 3.9);
\draw [line width=1, \tblue] (0.0, \hwfrac 4.1) -- (0.099, \hwfrac 4.12) -- (1.25, \hwfrac 5.00);
\draw [line width=\minthick, dotted, \tblue] (1.25, \hwfrac 5.00) -- (1.75, \hwfrac 5.4) ;
\draw [line width=1, dashed] (2.5,0) -- (2.5,4) ;
\draw (3.2,3) node {$k=k_\textup{max}$} ;
\end{tikzpicture}
}}
\caption{Structural spatial dispersion 
 in a Bragg mirror.
The periodic structure gives it similarities
 with the torus case of fig. \ref{fig-geom-torus}, 
 but in the maximum and minumum $k$ regions, 
 the two branches bend apart to create bandgaps, 
 where certain $(\omega,k)$ combinations
 do not propagate.
The dispersion relation is not linear, 
 does include the origin, 
 and is not continuous.}
\label{fig-struct-bragg}
\end{figure}
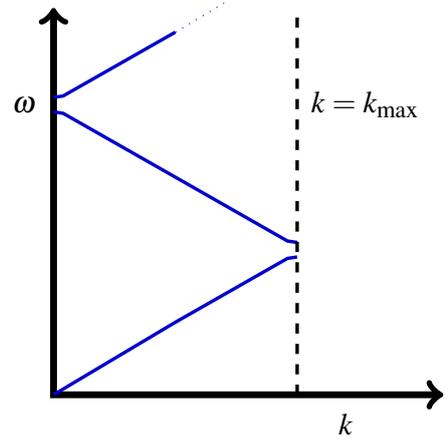

For our purposes it is sufficient to note that in cases where  $|z|>1$, 
 $\kappa$ becomes complex and $|\lambda|$ is no longer unity.
These represent diverging (non-periodic) solutions
 which are physically prohibited
 and so do not form part of the dispersion relation.
The resulting dispersion relation is nonlinear, 
 and is discontinuous across the (reflective) bandgap, 
 although continuous elsewhere; 
 it is depicted in fig. \ref{fig-struct-bragg}.

%
\subsection{Wire media}\label{S-Structural-wire}

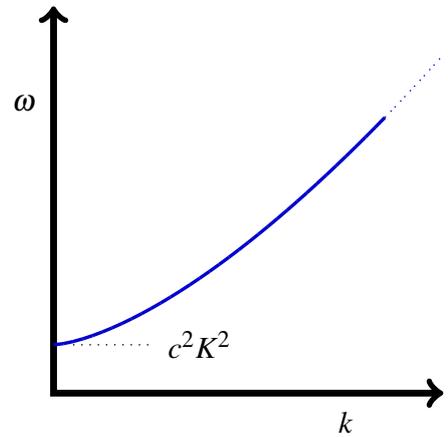
\begin{figure}
{\centering
\resizebox{0.70\columnwidth}{!}{
\begin{tikzpicture}
\draw [line width=2, <->] (0,4) -- (0,0) -- (4,0) ;
\draw (-0.3,3) node {$\pFrang$} ;
\draw [dotted] (0.2,0.5) -- (1,0.5) (1.5,0.5) node {$c^2 K^2$} ;
\draw (3,-0.3) node {$\pWaveVectorS$} ;
\foreach \i in {0.1,0.2, ..., 4.0} do {
 \draw [dotted, \tblue] ( {\i-0.1}, {\hwfrac 0.5*sqrt{(\i-0.1)*(\i-0.1)+0.50}} )
                       --
                      ( \i,      {\hwfrac 0.5*sqrt{\i*\i+0.50}} ) ;
} ;
\foreach \i in {0.1,0.2, ..., 3.4} do {
 \draw [line width=1, \tblue] ( {\i-0.1}, {\hwfrac 0.5*sqrt{(\i-0.1)*(\i-0.1)+0.50}} )
                       --
                      ( \i,      {\hwfrac 0.5*sqrt{\i*\i+0.50}} ) ;
} ;
\end{tikzpicture}
}}
\caption{Spatial dispersion in an approximated wire medium.
The cut-off $c^2 K^2$ gives a minimum allowed frequency, 
 but in the limit as $\pWaveVectorS$ becomes large 
 the curve approaches 
 the line $\pFrang = \beta \pWaveVectorS$.}
\label{fig-structural-wire}
\end{figure}

Wire media are a class of metamaterials consisting of
 a regular (rectangular) array of parallel wires or rods.
Thus the system is periodic along both the transverse axes,
 and (usually) uniform along the longitudinal one.
This is a hard problem to solve in the general case, 
 but results can be found when the wires have a 
 radius that is small compared to their spacing
 \cite{Belov-TV-2002jewa,Gratus-M-2015jo,Boyd-GKL-2018oe-tbwire}.
Conveniently, 
 it turns out that under approximation, 
 such media have a quadratic spatio-temporal dispersion relation.
This behaviour for the light in such a structure
 mimics that for light propagating in a plasma.
The approximate spatial dispersion relation is 
~
\begin{align}
\pFrang^2 - \beta^2 \pWaveVectorS^2 = c^2 K^2 
,
\end{align}
 which has a cut-off frequency  $c^2 K^2$.
The resulting dispersion curve is depicted on fig. \ref{fig-structural-wire}.

This is an example where the effect of inhomogeneity
 under approximation 
 induce a structural spatial dispersion 
 that mimics a mass-like term $\propto K^2$ in the 
 dispersion
 (see e.g. \cite{Kinsler-2013arxiv-kg2schro}).
After back Fourier transforming 
 this relation for the electric field $\pEfield$
 from $\pFrang,\pWaveVectorS$ into $\pTime,\pSpaceX$
 we get the wave equation
~
\begin{align}
  \pPderivT^2 \pEfield
 -
  \beta^2 
  \pPderivX^2
  \pEfield
 +
  c^2 K^2 \pEfield
&=
  0.
\end{align}
It is interesting to compare this wave behaviour, 
 and dispersion, 
 to the plasmon  dynamic case discussed in the next section: 
 despite their very different origins, 
 the qualitative behaviour is the same.

%
\section{Dynamic spatial dispersion}\label{S-Dynamic}

Dynamic spatial dispersion is a result of 
 non-trivial
 propagation properties of the material, 
 and how that affects the wave of interest that propagates through it.
For the spatial dispersion induced in wire media,
 we have already seen above
 that spatial properties can alter the effective dynamics
 of the light propagation.
Partly because of this, 
 and partly because of existing usage, 
 I categorise such propagation effects as spatial mechanism here because 
 it results from excitations (spatially) moving
 through the medium\footnote{Note that
  it is arguable that it should be instead regarded as 
  a space-time effect,
  due to its wave-like underlying model.
 However, 
  here I will retain it as a pure spatial effect 
  for consistency with existing usage;
  also note that the time/frequency-like $\pFrang^2$ term is not modified
  from that in the usual vacuum case.}, 
 {and so that speed is therefore the property which
 makes the phenomenon a distinctly spatial one.}

One might consider quite a wide variety of wave models
 when treating dynamic spatial dispersion; 
 for example, 
 in acoustic or elastic media, 
 we might think that any of the 
 somewhat eclectic selection in \cite{Kinsler-2018jpco-fbacou}
 could be suitable.
Indeed, 
 if used as a priori mechanisms,
 they might be, 
 but it is worth noting that such wave models result
 from applying approximations 
 to an underlying, 
 more complicated microscopic description.
As such, 
 the physical effects treated are 
 more like side-effects of
 structural spatial dispersion as treated in section \ref{S-Structural}, 
 rather than native dynamic spatial dispersion.

Because of such complications, 
 here I will consider 
 {just two cases of dynamic spatial dispersion,
 namely coupled waves and}
 the hydrodynamic model for plasmonics (HMP)
 \cite{Ciraci-PS-2013cphc}. 
{Notably, 
 the HMP is}
 a simple wave equation, 
 which is (only and exactly) second order 
 in both space and time it avoids issues of causality 
 as long as its built-in speed parameter
 stays less than that of light.
The origin of the resulting dispersion is 
 a direct result of the material properties,
 where the propagating wave of interests drives
 excitations in the medium, 
 and those excitations then affect the propagation of the original wave.

%
\subsection{Coupled waves}\label{S-Dynamic-coupled}

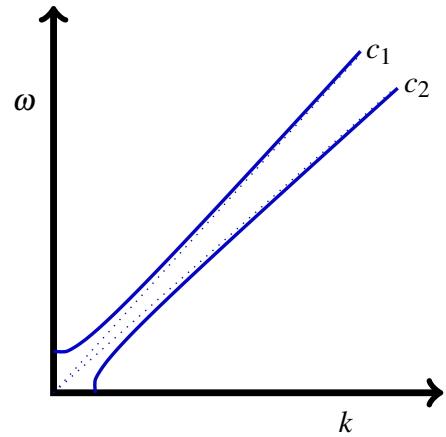
\begin{figure}
\def\rrr{0.9}
{\centering
\resizebox{0.70\columnwidth}{!}{
\begin{tikzpicture}[domain=0:3.5,xscale=1,yscale=1]
\draw [line width=2, <->] (0,4) -- (0,0) -- (4,0) ;
\draw (-0.3,3) node {$\pFrang$} ;
\draw (3,-0.3) node {$\pWaveVectorS$} ;
\draw [dotted, \tblue] (0.0,0.0) -- ({\rrr*3.5},3.5) node [black] {~\quad$c_1$};
\draw [dotted, \tblue] (0.0,0.0) -- (3.5,{\rrr*3.5}) node [black] {~~\quad$c_2$} ;
\draw[line width=1, smooth, \tblue] plot ({\rrr*\x}, {\x + 0.12/(\x+0.28)}) ;
\draw[line width=1, smooth, \tblue] plot ({\x + 0.12/(\x+0.28)},{\rrr*\x} ) ;
\end{tikzpicture}
}}
\caption{Spatial dispersion in a coupled wave system, 
 where $c_1>c_2$.
The dotted lines indicate the uncoupled dispersionless case(s), 
 whilst the solid lines indicate the spatial dispersion.
The intercepts are $\pFrang^4 = u_1 u_2$
 and $\pWaveVectorS^4 = u_1 u_2 / c_1^2 c_2^2$.
}
\label{fig-dynamic-coupled}
\end{figure}

{Perhaps the simplest dynamic case
 results from two coupled dispersionless waves 
 with different wave speeds.
Since, 
 as above, 
 we consider speed a spatial property, 
 this can give rise to spatial dispersion.
One might imagine, 
 perhaps, 
 a birefringent crystal 
 engineered to weakly couple the two field polarizations together, 
 where the reason for the speed difference is the 
 anisotropic properties of the crystal
 and the different field orientations.
The two coupled wave equations for fields $E_1$, $E_2$
 with speeds $c_1$, $c_2$, 
 are 
~
\begin{align}
  \pPderivT^2 \pEfield_1
 -
  \pPderivX {c_1^2} \pPderivX \pEfield_1
&=
  u_1
  \pEfield_2
,\\
  \pPderivT^2 \pEfield_2
 -
  \pPderivX {c_2^2} \pPderivX \pEfield_2
&=
  u_2
  \pEfield_1
,
\end{align}
 where $u_1$ and $u_2$ are the coupling parameters, 
 which here are taken to both be non-zero.  
When solved, 
 we see that the dispersion relation for the combined system is
~
\begin{align}
  \left(
    \pFrang^2 
   -
    c_1^2 \pWaveVectorS^2
  \right)
  \left(
    \pFrang^2 
   -
    c_2^2 \pWaveVectorS^2
  \right)
&=
  u_1 u_2 
.
\end{align}
This quartic expression, 
 whose solutions 
 are shown schematically on figure \ref{fig-dynamic-coupled}, 
 indicates that 
 the resulting system is spatially dispersive. 
}

%
\subsection{Plasmons}\label{S-Dynamic-plasmons}

\def\Pfield{\pXemXpolarizationv}
\def\wPlasma{\omega_{\textup{P}}}
\def\kPlasma{k_{\textup{P}}}

\def\Qfield{{\Psi}}   
\def\qfield{{\psi}}   
\def\ppphi{\pXemXspotential} 

\def\plloss{\gamma}          
\def\plspeed{\beta}          
\def\pldriving{\alpha}       
\def\plratio{\sigma}         

\def\Jfield{\pCurrentv}
\def\Nfield{N}               
\def\nfield{\eta}            

The HMP
 has recently found widespread popularity in the field of plasmonics
 and electromagnetism.
One of its key features is that
 unlike simpler plasmonics approaches based on the 
 Drude model for the dielectric permittivity, 
 it incorporates spatial derivatives terms 
 which represent the dynamics of the charge distribution.
Although these are often called ``non-local'' effects, 
 they are more usefully called propagation effects, 
 since they are not non-local in any sense that 
 violates relativistic (signalling) constraints on the physics.
However, 
 although much of its usage in plasmonics is recent, 
 the basic model itself dates back to the 1970's \cite{Eguiluz-Q-1976prb}
 and has been used in a number of other contexts 
 \cite{Fedorov-BRG-2005prb,Baranov-FRM-2003prb}.

One point regarding naming conventions that needs careful attention here
 is the use of ``plasmon'',
 which can be applied in various ways, 
 just as ``polariton'' can refer to a range of phenomena.
Here I use plasmon to mean the medium excitation only,
 i.e. the propagating disturbance in the electron gas
 (or dielectric polarization); 
 I do not mean the coupled electromagnetic -- electron-gas system
 which exhibits spatial dispersion\footnote{I would regard 
  this coupled system as a kind of polariton;
 i.e. a plasmon-polariton}.

The equation for the HMP in a 1D case is 
 \cite{Ciraci-PS-2013cphc}
~
\begin{align}
  \pPderivT^2 \Pfield
 +
  \plloss \pPderivT \Pfield
 -
  \pPderivX \plspeed^2 \pPderivX \Pfield
&=
  \pPermittivityVac
  \wPlasma^2
  \pEfieldv
,
\label{eqn-CPShydrodynamicmodel}
\end{align}
where 
 the polarization field is $\Pfield \equiv \Pfield(\pTime,\pSpaceX)$ 
 and
 the driving electric field is $\pEfieldv \equiv \pEfieldv (\pTime,\pSpaceX)$.
The speed of disturbances in the polarization field $\pEfieldv$ is $\plspeed$,
 polarization losses are given by  $\plloss$,
 and $\wPlasma^2$ is a measure of how strongly the electric field
 drives disturbances in the amplitude of $\Pfield$.
With respect to discussions about the utility and meaning of the HMP,
 we might consider \cite{Kinsler-2018arxiv-dehydro}
 either the deconstructed plasmon model
 or the variant current-based version,
 as opposed to the usual dielectric polarization version as used here.

Taking the 1D case for simplicity, 
 then after a double transform into $\pFrang$ and $\pWaveVectorS$,
 while also ignoring losses\footnote{Losses are ignored
  because they are solely a temporal effect, 
  and these examples are intended to present purely spatial mechanisms.}
 we get
~
\begin{align}
  \pFrang^2 \Pfield
 -
  \pWaveVectorS \plspeed^2 \pWaveVectorS \Pfield
&=
  \pPermittivityVac
  \wPlasma^2
  \pEfieldv
.
\label{eqn-CPShydrodynamicmodel-w}
\end{align}
When linked to the standard EM wave equation
 we find that this gives a homogeneous effective permittivity of
~
\begin{align}
  \pPermittivity(\pFrang,\pWaveVectorS)
&=
  \pPermittivityVac
  \left(
    \frac{\wPlasma^2}
         {\pFrang^2 
          - \pWaveVectorS^2 \plspeed^2 }
  \right)
,
\end{align}
 and a dispersion relation
~
\begin{align}
  \pFrang^2
 -
  \plspeed^2 \pWaveVectorS^2
 -
  \wPlasma^2
 &=
  0
.
\label{eqn-CPShydrodynamicmodel-dispersion}
\end{align}

If rewritten as
 a wave equation for light--plasmon polaritons, 
 and with the substitution $\wPlasma = c \kPlasma$,
 this is
~
\begin{align}
  \pPderivT^2
  E
 - 
  \plspeed^2 \pPderivZ^2
  E
 +
  c^2 \kPlasma^2
  E
&=
  0
.
\end{align}
This polariton equation can be factorized into two first order pieces
 \cite{Kinsler-2018arxiv-dehydro} in a way which emphasizes
 a spatial origin for the behaviour.
In the freely propagating case without driving terms,
 and with the auxilliary field $Q$, 
 this is 
~
\begin{align}
  \pPderivT E
&=
  \left( \plspeed \pPderivZ + c \kPlasma \right)
  Q
\\
  \pPderivT Q
&=
  \left( \plspeed \pPderivZ - c \kPlasma \right)
  E
.
\end{align}

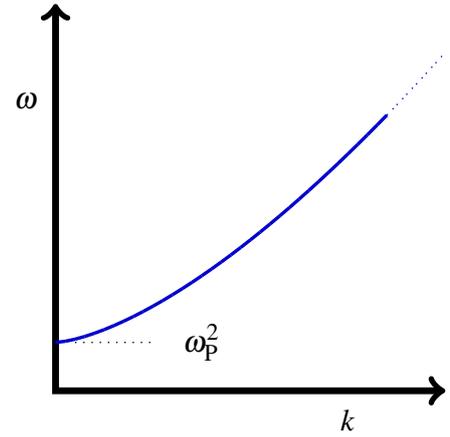
\begin{figure}
{\centering
\resizebox{0.70\columnwidth}{!}{
\begin{tikzpicture}
\draw [line width=2, <->] (0,4) -- (0,0) -- (4,0) ;
\draw (-0.3,3) node {$\pFrang$} ;
\draw [dotted] (0.2,0.5) -- (1,0.5) (1.5,0.5) node {$\wPlasma^2$} ;
\draw (3,-0.3) node {$\pWaveVectorS$} ;
\foreach \i in {0.1,0.2, ..., 4.0} do {
 \draw [dotted, \tblue] ( {\i-0.1}, {\hwfrac 0.5*sqrt{(\i-0.1)*(\i-0.1)+0.50}} )
                       --
                      ( \i,      {\hwfrac 0.5*sqrt{\i*\i+0.50}} ) ;
} ;
\foreach \i in {0.1,0.2, ..., 3.4} do {
 \draw [line width=1, \tblue] ( {\i-0.1}, {\hwfrac 0.5*sqrt{(\i-0.1)*(\i-0.1)+0.50}} )
                       --
                      ( \i,      {\hwfrac 0.5*sqrt{\i*\i+0.50}} ) ;
} ;
\end{tikzpicture}
}}
\caption{Dynamic spatial dispersion -- 
 a typical dispersion curve based on 
 how the EM field couples to 
 the hydrodynamic model for plasmons.
The cut off frequency $\wPlasma$ gives a minimum allowed frequency, 
 but in the limit as $\pWaveVectorS$ becomes large 
 the curve approaches 
 the line $\pFrang = \plspeed \pWaveVectorS$.}
\label{fig-dynamic}
\end{figure}

The behaviour of this polariton model
 can be seen on figure \ref{fig-dynamic}, 
 and
 shows the dynamic spatial dispersion, 
 with the dispersion relation 
 being non-linear and lacking a $(0,0)$ solution.

\section{Discussion: Spatial dispersion in context}\label{S-discussion}

Material properties generally are a result of 
 both spatial (structural) and temporal effects, 
 and often these are coupled together.
Even apparently simple properties such as the refractive index
 (which is usually simply a proxy for the material permittivity)
 are in fact a complicated combination  
 of the (many) individual temporal responses
 of the material's constituent atoms and molecules.
In a crystal, 
 the electronic structure of each atom 
 gives that atom its own independent temporal response, 
 and the arrangement of those atoms in space
 specifies the material's spatial (structural) properties. 
The combination gives the crystal a combined spatio-temporal dispersion
 which has a quite complicated behaviour.

{One important description that covers both temporal and spatial dispersion
 is based on a Fourier re-representation 
 of the medium reponse --
 i.e. the polarization or magnetization --
 as determined by integrals over space and time
 of a response kernel $K$ and field product.
A material polarization, 
 i.e. its response to an electromagnetic field
 can be given by such a convolution in 1+1D as 
\begin{align}
  P(x,t)
&=
  \int_{-\infty}^t dt'
  \int_{-\infty}^{\infty} dx
    ~~
    K(x,x',t,'t)
    E(x',t')
.
\label{eqn-superconvolution}
\end{align}
Note, 
 however, 
 that this is a very general expression,
 and that solely temporal or solely spatial dispersion
 are distinct and limited subsets
 of the wide range of spatiotemporal responses it can describe
 \footnote{\cBlue{It is important to note that exisiting 
usages of the term ``spatial dispersion'' include ones
where a material model of the type \eqref{eqn-superconvolution}
is an \emph{essential} feature; and any reading of the existing 
literature needs to be understood in that context.
Under such an existing usage, 
 the material properties of (e.g.) an optical fibre
 (even one represented only by spatial inhomogeneity
  and without any time dependence at all, 
  i.e. it \emph{only} has spatial properties)
  would not require description by \eqref{eqn-superconvolution}, 
  and so the dispersive properties it exhibits would not
  be classified as spatial dispersion.
 In contrast,
  and according to the argument laid out in the Introduction, 
  in Sec. \ref{S-Structural} 
  I do classify such dispersive properties
  as (structural) spatial dispersion.}}.
}

{These spatiotemporal complications
 were deliberately not addressed in the preceeding sections,
 because that would have obscured the goal of defining
 those specifically spatial features which by themselves can induce dispersive behaviour.
Indeed when we talk of the origins of dispersion, 
 whether temporal or spatial, 
 it is important to recognise that those origins
 (or their categorization)
 is really with respect to our
 models or representations of the system, 
 and not with respect to its exact physical properties}.

{Generally, 
 in practical cases there are many fine details can be ignored.
If we are happy with the approximation that 
 ``the refractive index of window glass is about 1.5'', 
 then we have ignored its true spatio-temporal dispersion, 
 and any residual solely-spatial or solely-temporal dispersion, 
 and settled for describing it in a non-dispersive way.
Given a regime of such (effectively) non-dispersive media, 
 we can then add large-scale (i.e. non-microscopic) spatial variation, 
 with e.g. two or more types of non-dispersive media, 
 and look at solely spatially dispersive effects, 
 as I have done above.}

However, 
 if we want to reproduce how light of different colours
 travels at different speeds in glass, 
 then we must retain the temporally dispersive part of the glass behavior.
Further,
 if we want to show
 how a glass prism can split white light into a rainbow,
 we also add in some of the spatial information, 
 i.e. only the prism's shape, 
 since describing the particular spatial arrangement of the atoms and molecules
 within the glass is superfluous.
In such natural media it is typically the case
 that the scales on which microscopic and bulk properties act
 are distinct enough so that we can treat them additively.

In contrast to natural media, 
 metamaterials tend to have dispersive behaviour 
 which is harder to approximate.
There are two significant reasons for this.
Firstly, 
 metamaterial design often relies on time dependent (dynamic) behavior
 to get strong responses, 
 so since the impinging wave field is --
 by design --
 rather near the metamaterial resonance, 
 and we cannot easily ignore the temporal contributions to dispersion.
Secondly,
 because metamaterials are themselves constructed of natural materials, 
 their unit cell scales are larger with respect to the impinging wavelengths
 than for natural materials where unit cells are atomic of molecular in size;
 thus we cannot easily ignore the spatial (structural)
 contributions to dispersion.
Indeed, 
 if we want to engineer a magnetic response, 
 then we cannot make a metamaterial cell size
 ``negligible'' in any useful sense \cite{Merlin-2009pnas}.

An important point to emphasize is that many types of metamaterials
 rely on both their spatial and temporal properties to work.
For example,
 many (such as the split ring resonator (SRR))
 are based on shaped metal structures, 
 in which electric currents can be induced
 by any impinging electromagnetic fields.
Once induced, 
 such currents can then follow their own shape-dependent
 time-domain dynamics,
 whilst also being driven by, 
 and emitting into,
 the field.
Strictly, 
 therefore, 
 the true response of such metamaterials
 is due to interlinked spatial and temporal dispersions.
However, 
 in the appropriate (or convenient) limits, 
 the spatial properties can often be homogenized away,
 leaving only the temporally dispersive properties as significant.

The extra work required to reduce a lattice of metamaterial cells
 is significant,
 and has lead to a great interest in how to measure, 
 simplify, 
 or otherwise describe their dispersive properties --
 leding to the topic known as \emph{homogenization}
 \cite{HomogMetaMat-2013pnfa}.
This goal of getting approximate 
 but still sufficiently accurate dispersion relations
 for uniform arrays of unit-cell structures
 has been widely examined, 
 but except in rare cases \cite{Belov-TV-2002jewa,Mnasri-KSPR-2018prb}
 the spatial properties are either
 ignored (as in e.g. the F-model for split-ring resonators (SRRs)
  \cite{Pendry-HRS-1999ieeemtt,Kinsler-2011ejp})
 or poorly resolved.
It is well beyond the intended scope of this paper
 to attempt to present the wide variety of possible spatial dispersia
 generated by metamaterials, 
 but the interested reader might perhaps
 look at either the Special Issue of PNFA \cite{HomogMetaMat-2013pnfa}, 
 or the more recent posibilities covered by the work of 
 Mnasri et al. \cite{Mnasri-KSPR-2018prb} and 
 Khrabustovskyi et al. \cite{Khrabustovskyi-MPSR-2017arxiv}.

{As a final point, 
 the plasma-like dispersion relations
 in \ref{S-Structural-wire} and \ref{S-Dynamic-plasmons}, 
 have the same quadratic form as for a Drude model.
However, 
 the wire medium and the plasmon model result from spatial properties
 and so suffer spatial dispersion, 
 and this contrasts with the Drude model whose dispersion 
 results from a temporal response, 
 and is thus temporally dispersive.
This is not in contradiction -- 
 instead this emphasises that any attribution
 of a ``spatial'' or ``temporal''  motivations
 to a particular dispersion relation must depend on 
 some underlying piece of physics that we have used in our model
 of the system.}

%
\section{Summary}\label{S-conclusion}

Here I have addressed the basic causes of spatial dispersion, 
 and briefly summarized examples of each.
However, 
 since 
 the term ``spatial dispersion'' is often used rather loosely,
 I first set out to specify clearly 
 what I meant by dispersion, 
 and what kinds of \emph{spatial} properties might generate it: 
 notably by the
 geometric, 
 structural, 
 or dynamic properties of a system.
Since my classification is not based on existing
 (and often somewhat ad hoc)
 naming conventions or justifications for 
 phenomena called (or attributed to) ``spatial dispersion'',
 it may be challenging to adapt to or accept.
However, 
 it is a definition derived solely based on
trying to answer the question:
  What different types of (solely) spatial properties can
  result in dispersive behaviour?

%
\section*{Acknowledgments}

I am grateful for the support provided by
 STFC (the Cockcroft Institute ST/P002056/1)
 and EPSRC (Alpha-X project EP/N028694/1); 
 and more recently
 by the UK National Quantum Hub for Imaging (QUANTIC, EP/T00097X/1);
 and for the hospitality of Imperial College London.
{I also acknowledge the contribution of the anonymous referees for 
 the PNFA published version 
 \cite{Kinsler-2021pnfa-spatype};
 in particular 
 the first referee's comparing the Drude model dispersion 
 with that of wire media, 
 and 
 the second referee's suggestion 
 of the coupled wave system now discussed in section \ref{S-Dynamic-coupled}.}

%

\end{document}